\documentclass[prl,aps,twocolumn,superscriptaddress,showpacs]{revtex4}

\usepackage[T1]{fontenc}
\usepackage{ae}   
\usepackage{wrapfig}
\usepackage{float}
\usepackage{amsmath}
\usepackage{amssymb}
\usepackage{amsfonts}
\usepackage[pdftex]{graphicx}	
\usepackage{array}	
\usepackage{bm}		
\usepackage{dsfont}	
\usepackage{ulem}
\usepackage{color}
\usepackage{hyperref}
\usepackage{units}
\usepackage{gensymb}

\renewcommand{\vec}[1]{\textbf{\textit{#1}}}
\newcommand{\mx}[1]{\ensuremath{\left( \begin{matrix} #1 \end{matrix} \right) }}

\newcommand{\zref}[1]{(\ref{#1})}
\newcommand{\erw}[1]{\langle \, {#1} \, \rangle}
\newcommand{\tr}{\text{tr}}
\newcommand{\intinfty}{\int_{-\infty}^\infty}
\newcommand{\abs}[1]{\left|#1 \right|}
\renewcommand{\Im}{\text{Im}\,}
\newcommand{\meV}{\text{ meV}}

\begin{document}

\title{Coupling to real and virtual phonons in tunneling spectroscopy of
superconductors}

\author{Jasmin Jandke}

\affiliation{Physikalisches Institut, Karlsruher Institut f\"ur Technologie, 76131
Karlsruhe, Germany}

\author{Patrik Hlobil}

\affiliation{Institut f\"ur Theorie der Kondensierten Materie, Karlsruher Institut
f\"ur Technologie, 76131 Karlsruhe, Germany}

\author{Michael Schackert}

\affiliation{Physikalisches Institut, Karlsruher Institut f\"ur Technologie, 76131
Karlsruhe, Germany}

\author{Wulf Wulfhekel}

\affiliation{Physikalisches Institut, Karlsruher Institut f\"ur Technologie, 76131
Karlsruhe, Germany}

\author{J\"org Schmalian}

\affiliation{Institut f\"ur Theorie der Kondensierten Materie, Karlsruher Institut
f\"ur Technologie, 76131 Karlsruhe, Germany}

\affiliation{Institut f\"ur Festk\"orperphysik, Karlsruher Institut f\"ur Technologie,
76344 Karlsruhe, Germany}

\date{\today }
\begin{abstract}
Fine structures in the tunneling spectra of superconductors have been
widely used to identify fingerprints of the interaction responsible
for Cooper pairing. Here we show that for scanning tunneling microscopy
(STM) of Pb, the inclusion of inelastic tunneling processes is essential
for the proper interpretation of these fine structures. For STM the
usual McMillan inversion algorithm of tunneling spectra must therefore
be modified to include inelastic tunneling events, an insight that
is crucial for the identification of the pairing glue in conventional
and unconventional superconductors alike. 
\end{abstract}

\pacs{74.55.+v, 74.81.Bd, 74.25.Jb, 74.25.Kc}

\maketitle
Conventional superconductivity is caused by the attractive interaction
between electrons near the Fermi energy mediated by phonons \cite{BCS}.
This leads to the formation of a gap $2\Delta$ in the single particle
density of states (DOS) of the electrons, and to quasi-particle peaks
above and below the gap~\cite{Giaever60,Nicol60}. Eliashberg extended the
BCS theory to the limit of larger dimensionless electron-phonon coupling
constants $\lambda$, included a realistic electron-phonon coupling
and the detailed structure of the phonon spectrum~\cite{Eliashberg60}.
As a consequence, the quasi-particle peaks near the Fermi surface
are modified due to the interaction with phonons, leading to fine
structures in the electronic DOS near the peaks of the Eliashberg function $\alpha^{2}F(\omega)$
shifted by $\Delta$. $\alpha^{2}F(\omega)$
is the phonon DOS $F(\omega)$, weighted by the energy dependent electron-phonon coupling strength $\alpha^2(\omega)$.
These fine structures are due to the excitation of virtual phonons
(see Fig. 1). Experimentally, these fine structures in the electronic
DOS have been detected with electron tunneling spectroscopy on planar
junctions~\cite{Giaever62,Schrieffer63,Rowell62,Rowell63,RowellinParks,Giaever74,Suderow2002}.
In the pioneering work of McMillan and Rowell \cite{McMillan65},
the Eliashberg function could be reconstructed from the superconducting
DOS by an inversion algorithm taking into account the interaction
of electrons and virtual phonons. This method has been used to identify
fingerprints of the phononic pairing glue in the electronic spectrum
and thus to determine the pairing mechanism leading to superconductivity
\cite{Scalapino66,CarbotteReview}. It counts as a hallmark of condensed
matter physics.

An alternative way to determine the Eliashberg function is to measure
the energy dependence of the scattering of electrons with real phonons
in the normal state using inelastic tunneling spectroscopy (ITS) \cite{Leger69,Rowell69,Wattamanuik71,Klein73},
see Fig.1. This method is more direct, as the second derivative of
the tunneling current $I$ with respect to the bias voltage $U$ is,
under rather general assumptions, directly proportional to $\alpha^{2}F(\omega)$~\cite{Taylor91}.
Recently, this method has been combined with scanning tunneling microscopy
(STM) to obtain local information on the Eliashberg function of Pb
on a Cu(111) substrate~\cite{Schackert15}.

\begin{figure}
\centering \includegraphics[width=0.45\textwidth]{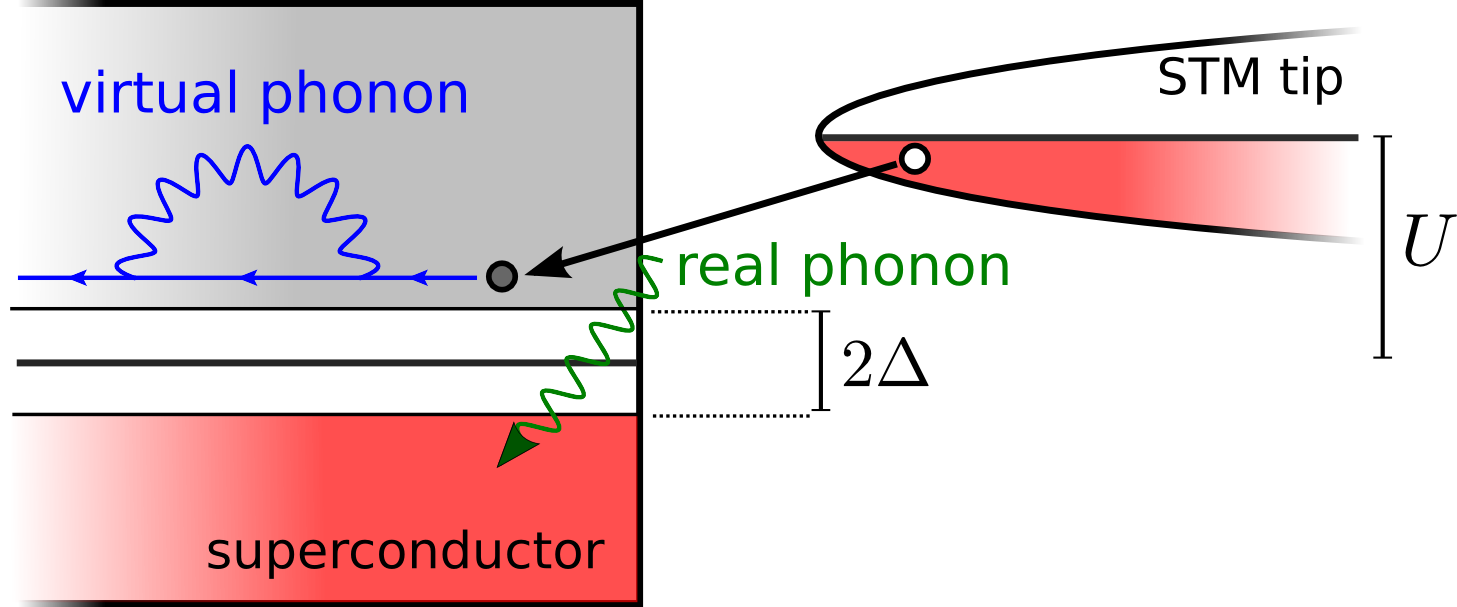}
\protect\caption{Illustration of the inelastic tunneling processes from a sharp tip
(right) into a superconductor (left) in real space. Filled states
are shown in red colour with energy along the vertical axis. The inelastic
tunneling process is accompanied by the excitation of real phonons
(green). }

\label{realvsvirtualphonons} 
\end{figure}

In this work, we determine experimentally and analyze theoretically
the tunneling conductance of Pb that is affected by the coupling to
real phonons via inelastic tunneling and virtual phonons via many-body
renormalizations. Comparing the two approaches to determine $\alpha^{2}F(\omega)$
on the same sample with the same tip of a low temperature STM, we
show that interpreting tunneling spectra of superconductors via the
McMillan inversion algorithm (and thus solely by its elastic contribution)
can be an incomplete description. We demonstrate that inelastic contributions
to the tunneling current can, in general, be of the same order as the elastic
contribution. We show that we can understand experimental STM data
from Pb tunneling in the normal and superconducting state, taking
into account both elastic and inelastic tunneling processes. The combined
analysis of elastic and inelastic tunneling processes is important
to correctly identify fingerprints of the relevant interactions in
the electronic DOS and to identify the pairing glue for superconductivity.
This is essential for conventional superconductors, such as Pb, but
is expected to be even more important for unconventional pairing states,
where an electronic pairing interaction is expected to fundamentally
change its character upon entering the superconducting state.

\begin{figure}[H]
\centering \includegraphics[width=0.4\textwidth]{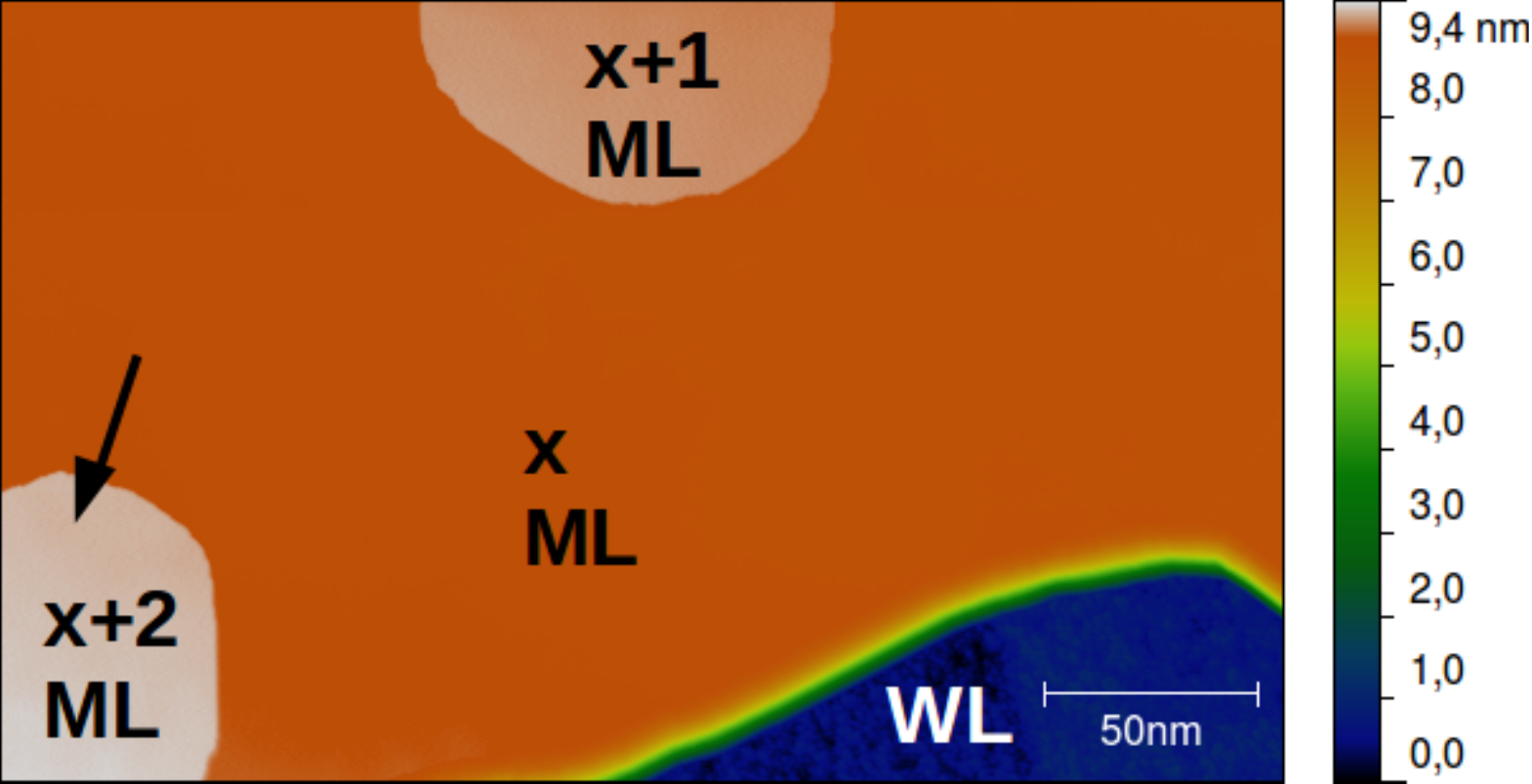} \protect\caption{STM topography of Pb on Si(111) ($300\times\unit[175]{nm^{2}},\ \unit[1]{V},\ \unit[1]{nA}$).
The thickness of the island was determined to $x\approx$ 30 monolayers.}

\label{fig:topo} 
\end{figure}

We start with experimental data for STM measurements on lead. Measurements were performed
with a home-build Joule-Thomson low-temperature STM (JT-STM)\cite{Lei1}
at temperatures about \unit[0.8]{K}. The JT-STM contains a
magnet which allows to suppress superconductivity. In order to ensure
that there is no significant inelastic signal of the tip at $\abs{U}<15\text{ mV}$,
we use a chemically etched tungsten tip, known to have a weak electron-phonon
coupling~\cite{McMillan1968tungsten}. The highly n-doped Si(111) crystals
were carefully degassed at \unit[700]{\degree C} for several
hours and then flashed to \unit[1150]{\degree C} for 30 seconds
to remove the native oxide. Lead was evaporated at room temperature
from a Knudsen cell with a nominal thickness of 19 monolayers (ML). After deposition the samples were immediately
transferred to the cryogenic STM. In agreement with previous studies
\cite{Brun,Eom2006,Altfeder1997}, flat-top, wedge-like islands of local thickness around 30ML were
observed (see Fig. \ref{fig:topo}), i.e. extended 3D islands appear on top
of a metallic wetting layer (WL) \footnote{Note that the extensions of the lead islands are typically larger than the 400 $\times$ 400 nm$^2$ STM images of the surfaces giving a minimal island size of  0.16 $\mu$m$^2$.}. 
The islands are Pb single crystals with their
$\langle111\rangle$ axis perpendicular to the substrate \cite{Eom2006,Weitering92,Jalochowski82}.
The first (second) derivative of the tunneling current $dI/dU$ ($d^{2}I/dU^{2}$)
of the islands was measured using a lock-in amplifier with a modulation
voltage of $U_{\text{mod}}=\unit[439]{\mu V}$.

\begin{figure}
\includegraphics[width=0.47\textwidth]{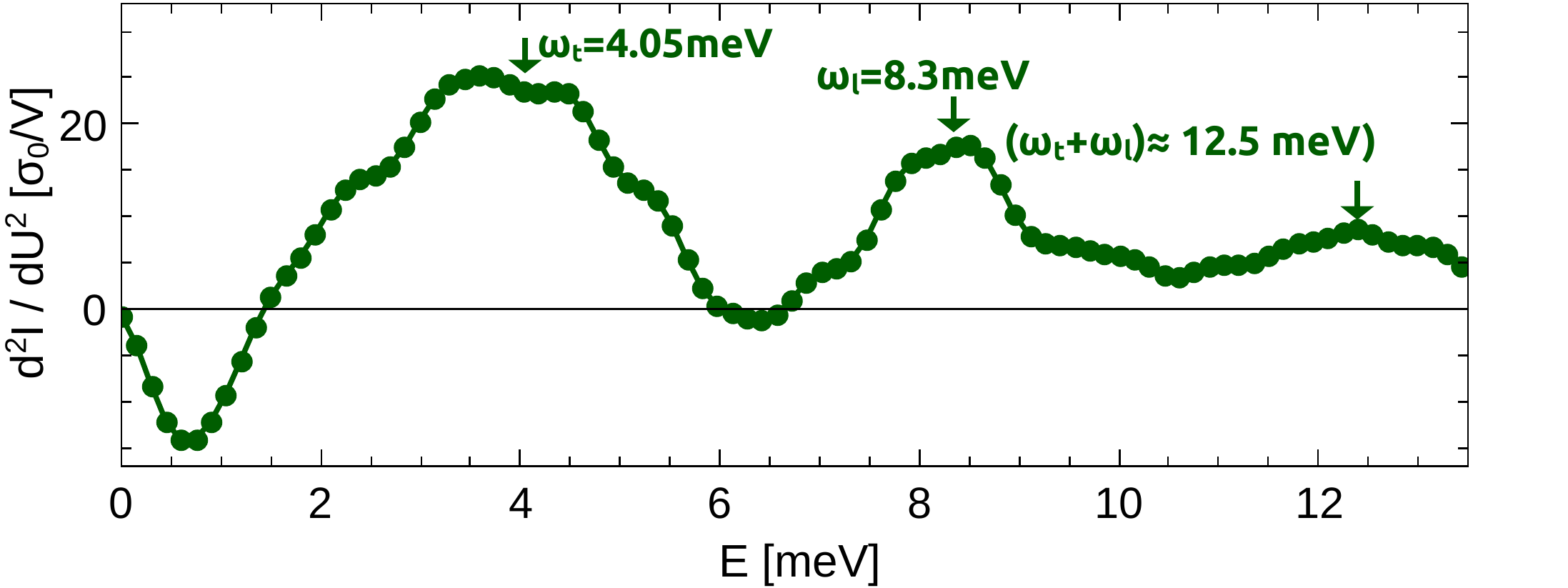} \protect\caption{Second derivative $d^{2}I/dU^{2}\sim\alpha^{2}F_{\text{tun}}(eU)$
measured in the normal conducting state ($T=0.8$ K, $B=1$ T).}
\label{nl2} 
\end{figure}

While the electrons in the $\approx30$ML Pb film on Si have quantized $k_{z}$ leading to the flat island growth,
the phonon DOS of the finite thickness films is rather similar to that of bulk Pb as indicated by first principle calculations~\cite{Heid2013,Heid2010}.
As a first measurement, we therefore determine $\alpha^{2}F_{\text{tun}}(\omega)$
of lead directly with ITS in the normal state. Pb islands were forced
to the normal state by applying a magnetic field of 1T normal to the
sample plane. 
Since the sample is in the normal state, no renormalization of the BCS density of states near the Fermi energy due to virtual phonons arises. Thus renormalization effects by virtual phonons can be neglected in $d^{2}I/dU^{2}$ and experimental features in $d^{2}I/dU^{2}$  correspond to inelastic tunneling. 
Fig.~\ref{nl2}
shows the measured $d^{2}I/dU^{2}$ spectrum clearly revealing the
two characteristic phonon peaks that are also seen in the Eliashberg
function $\alpha^{2}F(\omega)$ determined by Ref.~\cite{McMillan65}.
Below we show explicitly that in the normal state $d^{2}I/dU^{2}$
is proportional to $\alpha^{2}F_{\text{tun}}(e\, U)$. These peaks
at $U=$ \unit[4.05]{mV}$\approx\omega_{t}$ and $U\approx$
\unit[8.3]{mV}$\approx\omega_{l}$ (FWHM $\gamma_{t}=1.076$meV
and $\gamma_{l}=0.60$meV) coincide with the energies of the transversal
and longitudinal van Hove singularities
in the phonon DOS of lead \cite{Heid2010,Brockhouse62}. The additional
peak at $U\approx$ \unit[12.5]{mV} can be explained by tunneling
processes via two-phonon emission \footnote{Note that also the second peak may already include such two-phonon processes.}. 

The key implication from Fig.~\ref{nl2} for the superconducting
state is, however, that we must include inelastic contributions to the superconducting tunneling
spectrum in a consistent fashion. Before we present
our experimental data of the superconducting state, we summarize the
theoretical description of the tunneling conductance in the superconducting state
including inelastic contributions.

The Hamiltonian $\mathcal{H}=\mathcal{H}_{0}+\mathcal{H}_{\text{t}}$
used in our analysis of the combined substrate and tip consists of
free electrons in the tip and electrons interacting with phonons in
the substrate (we set $\hbar=1$): 
\begin{align}
\mathcal{H}_{0} & =\sum_{\vec{p},\sigma}\epsilon_{\vec{p}}^{T}c_{\vec{p},\sigma}^{\dagger}c_{\vec{p},\sigma}+\sum_{\vec{k},\sigma}\epsilon_{\vec{k}}^{S}c_{\vec{k},\sigma}^{\dagger}c_{\vec{k},\sigma}+\sum_{\vec{q},\mu}\omega_{\vec{q},\mu}a_{\vec{q},\mu}^{\dagger}a_{\vec{q},\mu}\nonumber \\
 & \hspace{3mm}+\frac{1}{\sqrt{V_{S}}}\sum_{{\vec{k},\vec{k}'\atop \sigma,\mu}}\alpha_{\vec{k}-\vec{k}',\mu}c_{\vec{k},\sigma}^{\dagger}c_{\vec{k}',\sigma}\phi_{\vec{k}-\vec{k}',\mu}\,.
\end{align}
Here, $\phi_{\vec{q},\mu}=a_{\vec{q},\mu}+a_{-\vec{q},\mu}^{\dagger}$
is proportional to the lattice displacement, where $a_{\vec{q},\mu}$
is the the phonon annihilation operator for momentum $\vec{q}$ and
phonon-branch $\mu$ and with dispersion $\omega_{\vec{q},\mu}$.
$c_{\vec{k}/\vec{p},\sigma}^{\dagger}$ are the electron annihilation
operators for the two subsystems: The tip (quasi-momentum $\vec{p}$,
dispersion $\epsilon_{\vec{p}}^{T}$ and volume $V_{T}$) and the
superconductor (quasi-momentum $\vec{k}$, dispersion $\epsilon_{\vec{k}}^{S}$
and volume $V_{S}$). For the latter we include the electron-phonon
coupling $\alpha_{\vec{k}-\vec{k}',\mu}$ that gives rise to superconductivity.
The electron-phonon interaction in the tip is assumed to be small.
In addition, the tunneling part of the Hamiltonian includes elastic
and inelastic tunneling processes \cite{Taylor91,BennethDuke68}: 
\begin{align}
\mathcal{H}_{\text{t}} & =\frac{1}{\sqrt{V_{T}V_{S}}}\sum_{{\vec{k},\vec{p}\atop \sigma}}T_{\vec{k},\vec{p}}c_{\vec{k},\sigma}^{\dagger}c_{\vec{p},\sigma}+\text{h.c.}\,,\label{th2}\\
T_{\vec{k},\vec{p}} & =T_{\vec{k},\vec{p}}^{\text{e}}+\frac{1}{\sqrt{V_{S}}}\sum_{\vec{q},\mu}T_{\vec{k},\vec{p},\vec{q},\mu}^{\text{i}}\alpha_{\vec{q},\mu}\phi_{\vec{q},\mu}+\mathcal{O}(\phi_{\vec{q},\mu}^{2})\,.\nonumber 
\end{align}
The first term of the tunneling amplitude $T_{\vec{k},\vec{p}}$ describes
the elastic tunneling part, the second term corresponds to electron
transitions via the emission/absorption of phonons, see Fig.~\ref{realvsvirtualphonons}.
It is proportional to the bulk electron-phonon coupling $\alpha_{\vec{q},\mu}$~\cite{Taylor91}.
There can also be processes with a higher number of phonons, which
will be discussed later.

In order to determine the tunneling current we assume that the DOS
of the tip is constant $\nu_{T}(\omega)\approx\nu_{T}^{0}$ and that
the tunneling amplitudes are independent of momenta and phonon branches
$T_{\vec{k},\vec{p}}^{\text{e}}=t^{\text{e}}$ and $T_{\vec{k},\vec{p},\vec{q},\mu}^{\text{i}}=t^{\text{i}}$,
which is a reasonable approximation for STM~\cite{Giamarchi2011}.
Then, to leading order in $t^{\text{e}}$, the differential conductance
gives the well known result~\cite{Bardeen61,Cohen62,MahanBook} 
\begin{align}
\sigma^{\text{e}}(U)=\frac{dI^{\text{e}}}{dU} & =-e\sigma_{0}\intinfty d\omega\, n_{F}'(\omega+eU)\tilde{\nu}_{S}(\omega).\label{th3}
\end{align}
In the limit that $T$ is smaller than the electronic energy scales,
the conductance is just proportional to the the normalized electron
DOS $\tilde{\nu}_{S}(\omega)=\nu_{S}(\omega)/\nu_{S}^{0}$, where
$\nu_{S}^{0}$ is the normal state DOS of the superconductor at the
Fermi level. The conductance constant is given by $\sigma_{0}=4\pi e^{2}\abs{t^{\text{e}}}^{2}\nu_{T}^{0}\nu_{S}^{0}$
and $n_{F}$ is the Fermi function. In
the normal state, $\tilde{\nu}_{S}(\omega)$ is essentially constant
for small applied voltages and the second derivative of the elastic
current vanishes, as discussed above. In the superconducting state, the opening of the
superconducting gap and the excitation of virtual phonons lead to
the mentioned fingerprints of superconductivity and the pairing glue
in the elastic tunneling spectrum. Below we determine these structures
from the solution of the nonlinear Eliashberg equations for given
$\alpha^{2}F(\omega)$ and compare with our STM experiments.

The inelastic contribution to the differential conductance $\sigma^{\text{i}}(U)=\frac{dI^{\text{i}}}{dU}$
due to the excitation of single real phonons is for $U>0$ given by
the convolution 
\begin{align}
\sigma^{\text{i}}(U)=\sigma_{0}\frac{\abs{t^{\text{i}}}^{2}}{\abs{t^{\text{e}} }^{2} \nu_S^0}\intinfty d\omega\,\alpha^{2}F_{\text{tun}}^{T}(eU+\omega)\tilde{\nu}_{S}(\omega)n_{F}(\omega)\,.\label{th5}
\end{align}
in the limit that the thermal phonons can be neglected $T\ll\omega_{D}$.
The function $\alpha^{2}F_{\text{tun}}^{T}(x)=-\intinfty dy\,\alpha^{2}F_{\text{tun}}(y)n_{F}'(y-x)$
is a thermally broadened version of the weighted phonon DOS $\alpha^{2}F_{\text{tun}}(\omega)=\frac{\nu_S^0}{V_{S} }\sum_{\vec{q},\mu}\abs{\alpha_{\vec{q},\mu}}^{2}\delta(\omega-\omega_{\vec{q},\mu})$
that is closely related to the Eliashberg function $\alpha^{2}F(\omega)=\frac{1}{\nu_{S}^{0}V_{S}^{2}}\sum_{\vec{k},\vec{k}',\mu}\abs{\alpha_{\vec{k}-\vec{k}',\mu}}^{2}\delta(\omega-\omega_{\vec{k}-\vec{k}',\mu})\delta(\epsilon_{\vec{k}}^{S})\delta(\epsilon_{\vec{k}'}^{S})$.
Both have similar features but can differ in fine-structure and amplitude.

\begin{figure}
\centering \includegraphics[width=0.49\textwidth]{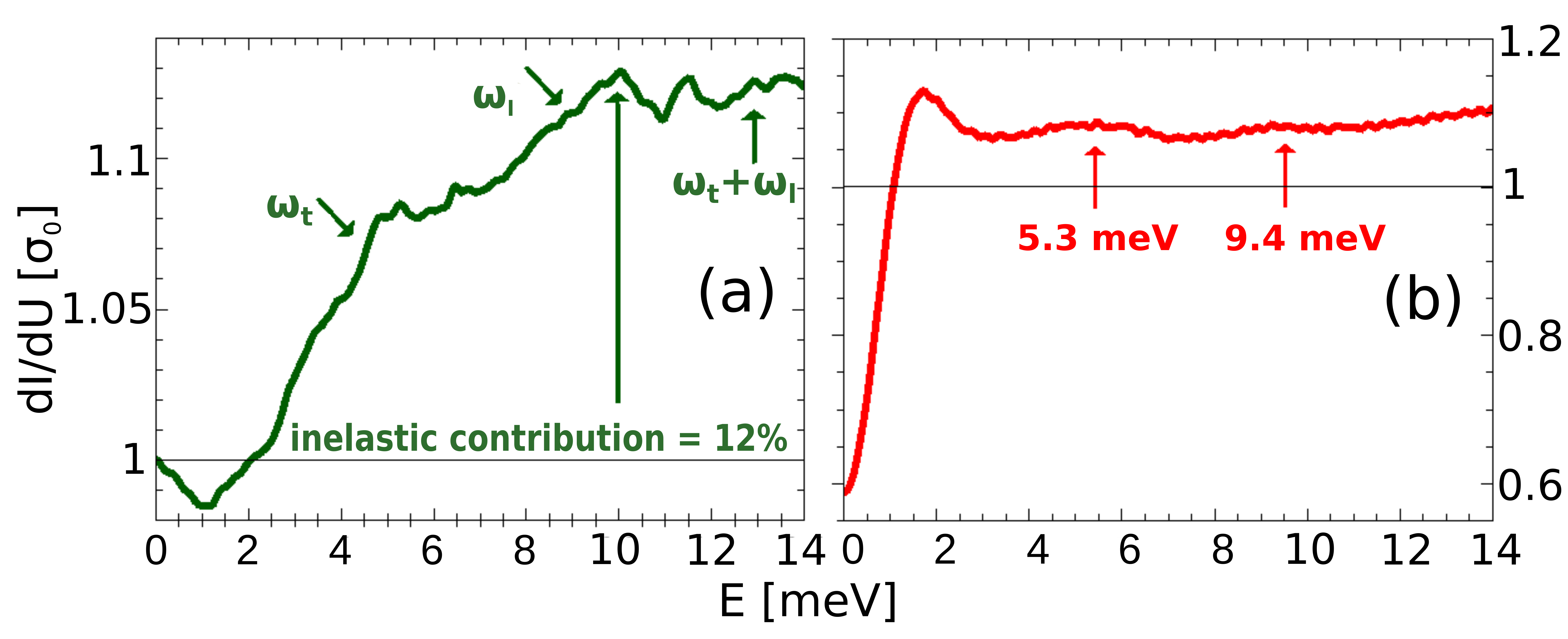} \protect\caption{Differential conductance $dI/dU$ in the normal (a) and superconducting
(b) state measured on the island
marked by arrow in Fig.~\ref{fig:topo}. The curves are normalized
to the zero bias conductance $\sigma(0)$ in the normal state. }
\label{dIdU} 
\end{figure}

The result~\zref{th5} is the generalization of the current in
the normal state, where $\frac{d^{2}I^{\text{i}}}{dU^{2}}\bigr|_{\text{NC}}\sim\text{sign}(U)\,\alpha^{2}F_{\text{tun}}(e\abs{U})$
is proportional to the weighted DOS of the phonons (or other collective
excitations of the system), see Ref.~\cite{BennethDuke68,Taylor91,KirtleyScalapino1990,Xiao1994}.
It naturally explains the results of Fig.~\ref{nl2} or the recent
STM measurements on Pb~\cite{Schackert15}. Our measurement further allows for an estimate of the inelastic tunneling amplitude $t^\text{i} \approx t^{\text e}/D$, which is inversely proportional to the characteristic energy scale of the off-shell electrons involved in the tunneling process. The normal state elastic  conductance $\sigma^{\text e}(U) \approx \sigma_0$ is not energy dependent for the applied biases $U$ and we emphasize that all spectra within this paper are normalized to $\sigma_0= \sigma(0)=\sigma^{\text e}(0)$ to point out the existence of inelastic tunneling contributions. The change in the conductance from 0 to 10 mV seen in Fig.~\ref{dIdU}a) is purely due to the inelastic tunneling. This leads to the condition $\sigma^{\text{i}}(10\, \text{mV}) \approx 12 \% \, \sigma_0 $, where $\sigma(0)=\sigma_0$ is the purely elastic contribution at zero bias.  Using the widely accepted Eliashberg function $\alpha^2 F(\omega)$ and the experimental DOS for lead~\cite{Gold60}, we can estimate for the characteristic off-shell electronic energy to be $D \approx 240\, \text{meV}$. Below, we will see that elastic and inelastic contributions to the fine-structure turn out to be comparable in magnitude.

In the superconducting state, the inelastic contribution Eq.\zref{th5}
has its major contribution slightly below the energy
of the phonon peaks shifted by the gap $\Delta$. Since inelastic
tunneling opens additional channels to the conductance, it will lead to positive contributions
to $d^{2}I/dU^{2}$ at positive bias. Elastic contributions are of opposite sign (see \zref{th3}).
Thus, pronounced peaks in the second derivative of
the tunneling current due to real phonons are followed by dips of same amplitude due to virtual phonons (for details see discussion of a single phonon mode in the Supplementary Material).
As we will see below, we find exactly these features in the tunneling current for the STM experiment on lead.

Tunneling processes with a higher number of excited phonons will give
similar terms as in \zref{th5} with higher convolutions of the Eliashberg-function
such as $\alpha^{4}F_{\text{tun}}^{2}(\omega)=\int d\omega'\alpha^{2}F_{\text{tun}}(\omega-\omega')\alpha^{2}F_{\text{tun}}(\omega')$
and one can formally absorb this contribution in a redefinition of
$\alpha^{2}F_{\text{tun}}$ (see Supplementary Material).

\begin{figure*}
	\includegraphics[width=0.75\textwidth]{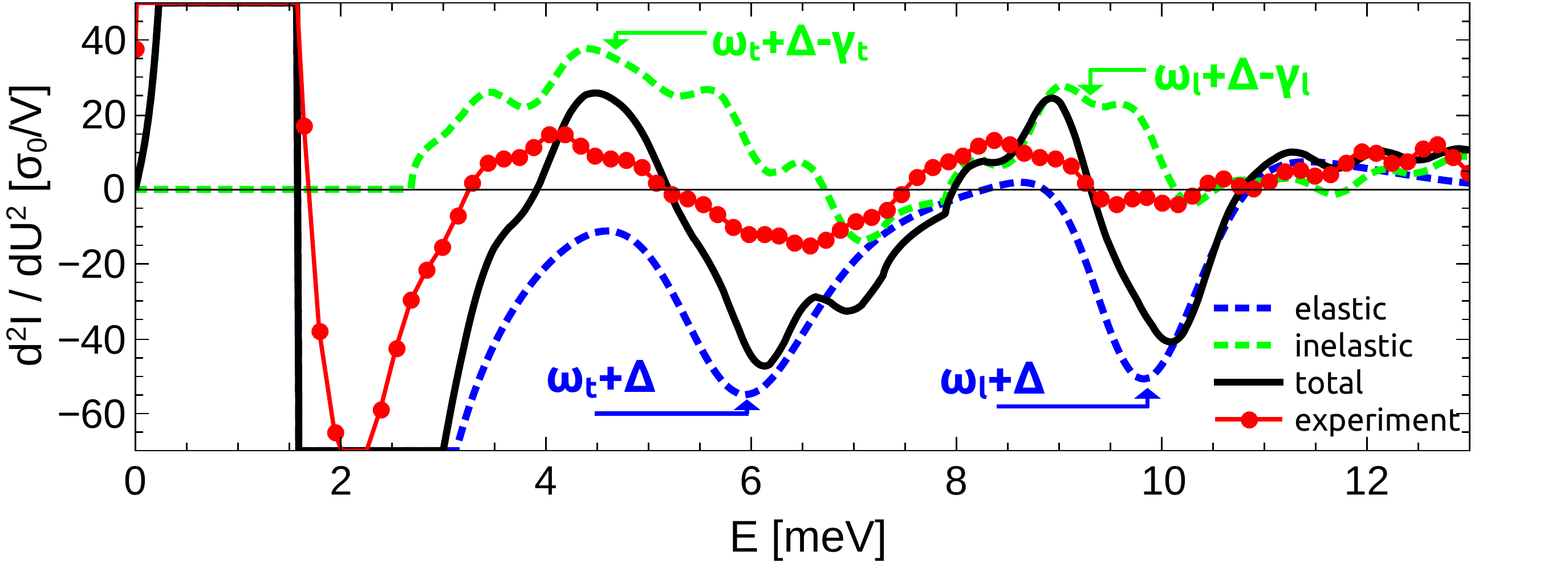} \protect\caption{Comparison of experimental data (red) and theoretical prediction in
		the superconducting state: Calculated elastic (blue), inelastic (green)
		and total (black) contribution to $d^{2}I/dU^{2}$ (the elastic current
		is convoluted with a Gaussian function with standard deviation $\sigma=0.31\meV$
		simulating the experimental broadening due to the modulation voltage
		of the lock-in technique). Characteristic peak-dip features around
		the zero axis can only be explained taking into account elastic and
		inelastic channels ($I^{\text{tot}}=I^{\text{el}}+I^{\text{inel}}$).}
	\label{sl2} 
\end{figure*}
Without magnetic field, the islands are in the superconducting state.
As the local thickness of the islands (30ML $\approx$ 10
nm) is significantly smaller than the bulk coherence length of lead (83 nm \cite{Kittel2005}),
the superconducting gap is not fully developed~\cite{Brun,Eom2006,Nishio2006,Nishio2008,Qin2009,Garcia2011,Bose2010},
which implies that the spectral weight of the coherence
peaks is accordingly smaller (see Fig.~\ref{dIdU}b)
). Besides the Bogoliubov quasiparticle peak one clearly observes
fine structures in the spectrum of the conductance around $U\approx$
\unit[5.3]{mV} and $U\approx$ \unit[9.4]{mV}. These
energies correspond to the van Hove singularities in the phonon DOS
$F(\omega)$ of lead shifted by the gap $\Delta\approx1.2\meV$ clearly
indicating electron-phonon interaction induced effects. Furthermore,
the typical $\omega/\sqrt{\omega^{2}-\Delta^{2}}$ behavior in the
BCS DOS is altered by the emergence of inelastic contributions at
biases $V_{0}>5mV$ . This is in contrast to previous measurements
on planar tunneling junctions of lead~ \cite{Giaever62,Schrieffer63,Rowell62,Rowell63,RowellinParks,Giaever74},
where these inelastic contributions were about one order of magnitude
smaller~\cite{Rowell69} than in our present experiment. The reason
is that inelastic tunneling events are enhanced in STM geometries,
if compared to planar tunneling junctions, because the momentum conservation
for momenta parrallel to the surface are less restrictive~\cite{Giamarchi2011}.\\
 Let us now investigate the second derivative of the tunneling current
in the superconducting state, which is significantly more sensitive
to the fine structure induced by the electron-phonon interaction.
For the theoretical spectrum, we first use a parameterization of the
$\alpha^{2}F(\omega),\mu^{*}$ from McMillan and Rowell~\cite{McMillan65,Galkin}
to solve the Eliashberg equations numerically~\cite{Schmalian96}
to obtain the lead DOS $\nu_{S}(\omega)$ in the superconducting state.
The elastic contribution to the second derivative is then easily calculated
using Eq.~\zref{th3}. For the inelastic contribution we use the
$\alpha^{2}F_{\text{tun}}(\omega)$ function (without the negative
dip at small voltages $U<2$ mV that comes from a zero bias anomaly)
and the calculated DOS $\nu_{S}(\omega)$ to determine the convolution
in Eq.~\zref{th5}, where the usage of the measured $\alpha^{2}F_{\text{tun}}(\omega)$
function automatically includes two-phonon processes and yields the
correct amplitude for the inelastic tunneling current. Note, that
the rapid wiggels on top of the calculated inelastic curve are
due to noise of the input data of the calculations, i.e. the experimental inelastic spectrum in the normal state. This noise is caused by residual mechanical vibrations on the level of \unit[300]{fm}. When convoluting the noisy experimental spectra with the DOS
for the calculation of the inelastic contribution in the superconducting
state, certain frequencies of the noise are amplified and show up
as small wiggels. Finally, we convoluted the elastic part 
\footnote{Note that we should not broaden the inelastic contribution as we use
the $\alpha^{2}F_{\text{tun}}(\omega)$ from the normal conductor
measurement that already includes broadening, see Supplementary Material
for details.} with a Gaussian distribution (standard deviation $\sigma=310\mu\text{eV}$
corresponding to an energy resolution of $744\mu\text{eV}$), describing
the experimental broadening due to the modulation voltage of the lock-in
detection~\cite{Klein73}. 

 In Fig.~\ref{sl2} we compare the experimental data with the theoretical
prediction of the elastic, inelastic and total contributions of the
second derivative of the current. The experimental data show 
peak-dip features around the zero axis at positions that correspond
to the characteristic longitudinal and transversal phonon energies
$\omega_{\text{t}/\text{l}}$ shifted by the gap $\Delta\approx1.2\meV$.
For both features there is a positive peak at $\approx\omega_{t,l}+\Delta-\gamma_{t,l}$
of the same magnitude as the corresponding dip at $\omega_{t,l}+\Delta$,
where $\gamma_{t,l}$ are the half-widths of the phonon peaks observed
in Fig.~\ref{nl2}. This is in contrast to the theoretical elastic
$d^{2}I^{\text{e}}/dU^{2}\sim\nu'(-eU)$ curve, which only shows the
typical dips around $\Delta+\omega_{\text{t/l}}$ predicted by the
Eliashberg theory. We note that conventional Eliashberg theory can
also have positive peaks, but the following dip will always be significantly
more pronounced (see also Fig. 4 in the Supplementary material). Therefore,
the observed peak-dip features cannot be explained by
pure elastic tunneling. However, the measured spectrum both in the
normal \textit{and} in the superconducting state can naturally be
explained when we combine inelastic and elastic contributions. As can be seen,
the total theoretical conductance $d^{2}I^{\text{tot}}/dU^{2}$ consisting
of elastic and inelastic channels fits the experimental
peak-dip features much better at the correct energies.

In summary, we demonstrated experimentally and theoretically that
in normal conducting Pb islands it is possible to directly measure
the collective bosonic excitation spectrum, here phonons, using STM.
In the normal conducting state, the obtained $d^{2}I/dU^{2}$ spectra
is proportional to the weighted phonon DOS $\alpha^{2}F_{\text{tun}}(\omega)$
and higher convolutions thereof. This is different in the superconducting
state of Pb. Here, the obtained second derivative $d^{2}I/dU^{2}=d^{2}I^{\text{e}}/dU^{2}+d^{2}I^{\text{i}}/dU^{2}$
spectra are a composition of elastic and inelastic tunneling processes
with fine structures in the same energy regime. While the elastic
part shows phonon features coming from self energy corrections (exchange
of virtual phonons) that appear mainly as
dips in the second derivative of the tunneling
current, the inelastic part shows features of $\alpha^{2}F_{\text{tun}}(\omega)$
shifted by the superconducting gap $\Delta$ giving rise to additional
peak features of the same amplitude at lower energies (excitation
of real phonons). A rather unique signature of these inelastic contributions
are peak-dip features in $d^{2}I/dU^{2}$ around
zero at $\Delta+\omega_{\text{ph}}$ in the superconducting state.
Those cannot be explained by only taking into account the elastic
part $d^{2}I^{\text{e}}/dU^{2}$. For this reason, the neglect of
inelastic processes in STM experiments %(cf. eg.~\cite{Shan2012,Niestemski2007,Chi2012,Song2014}) is
in general not justified. Hence, when analyzing STM tunneling spectra
via the McMillan inversion algorithm~\cite{McMillan65,Galkin}, that
gives the purely elastic contribution, one should carefully subtract
the inelastic contributions from the experimental tunneling current.
Otherwise grossly incorrect conclusions about the pairing glue would
be deduced from the tunneling spectrum.

Having found out experimentally and theoretically how elastic and
inelastic tunneling can be disentangled for STM in conventional superconductors,
the approach can be generalized to the investigation of corresponding
bosonic structures in high temperature superconductors such as cuprates
and iron pnictides in the future. A crucial difference to the phononic
pairing glue is that in case of electronic pairing, the bosonic spectrum
undergoes dramatic reorganization below $T_{c}$ in form of a sharp
resonance in the dynamic spin excitation spectrum~\cite{RossatMignod1991,Mook1993,Fong1999,Christianson2008,Inosov2010,Abanov2001}.
Our results imply that great care must be taken in the proper interpretation
of the tunneling spectra of these systems and that real and virtual
bosonic excitations must be disentangled in a fashion similar to our
analysis for lead.

\begin{acknowledgements}
The authors acknowledge funding by the DFG under the grant SCHM 1031/7-1 and WU 349/12-1.
	\end{acknowledgements}

\appendix

\section*{\Large Supplemental Material}

\section{Derivation of the tunnel current}
\subsection{Peturbative approach}    \label{secpertapproach}
The tunneling current is given by elementary charge times the change of the number of electrons $n_S=\sum_{\vec k,\sigma} c_{\vec k,\sigma}^\dagger c_{\vec k,\sigma}$  in the superconductor
\begin{align}
I &= - e \frac{d}{dt} \tr \bigl[ \rho(t) n_S \bigr]/\tr[\rho(t)] \nonumber \\
&= i e  \, \tr\bigl( \rho_0 \bigl[n_S(t),\mathcal H(t)  \bigr] \bigr) /\tr[\rho_0] \nonumber \\
&= i e \erw{\bigl[n_S(t),\mathcal H(t)  \bigr]}   \label{th4}
\end{align}
where $\rho(t)=U(t,\infty) \rho_0 U^\dagger(t,-\infty)$ is the time-dependent density matrix and $ \erw{\ldots}_0 = \erw{\rho_0 \ldots}= \erw{e^{-\beta \mathcal H} \ldots}$ is the expectation value of the system in thermal equilibrium with density matrix $\rho_0= e^{-\beta \mathcal H}$. A suitable formalism to calculate the current \zref{th4} is the Keldysh Green function method (we follow the notation of Ref.~[\onlinecite{Keldyshbook}]). The corresponding Keldysh action of the Hamiltonian (without bias voltage) employed in the main text of the paper is given by
{\small
	\begin{align}
	S&= S_0+S_{\text{t}}  \label{th5} \\
	S_0 &=  \int_{\mathcal C} dt \,  \sum_{\vec p,\sigma} \bar c_{\vec p,\sigma}(t) (i \partial_t - \epsilon_{\vec p}^T ) c_{\vec p,\sigma}(t)   \nonumber \\
	& \hspace{3mm}+ \int_{\mathcal C} dt \,  \sum_{\vec k,\sigma} \bar c_{\vec k,\sigma}(t) (i \partial_t - \epsilon_{\vec k}^S ) c_{\vec k,\sigma}(t)      \nonumber \\
	& \hspace{3mm}+ \int_{\mathcal C}dt \,   \sum_{\vec q,\mu} \bar a_{\vec q,\mu}(t) (i \partial_t - \omega_{\vec q,\mu} ) a_{\vec q,\mu}(t) \nonumber \\
	& \hspace{3mm}- \frac{1}{\sqrt{V_S}}\int_{\mathcal C} dt \,   \sum_{\vec k, \vec k'\atop \sigma,\mu} \alpha_{\vec k-\vec k',\mu} \bar c_{\vec k,\sigma}(t) c_{\vec k',\sigma}(t) \phi_{\vec k-\vec k',\mu}(t)    \nonumber \\
	S_{\text{t}}&= - \frac{1}{\sqrt{V_S V_T}}\int_{\mathcal C} dt \, \sum_{\vec k,\vec p} T_{\vec k,\vec p}^{\text{e}}   \bar  c_{\vec k,\sigma}(t) c_{\vec p,\sigma}(t)  \nonumber \\
	& \hspace{3mm} - \frac{1}{V_S \sqrt{V_T}}\int_{\mathcal C} dt \,  \sum_{\vec k, \vec p , \vec q \atop \sigma,\mu} T_{\vec k,\vec p,\vec q, \mu}^{\text{i}} \,  \alpha_{\vec q,\mu} \bar c_{\vec k,\sigma}(t) c_{\vec p,\sigma}(t) \phi_{\vec q,\mu} (t) \nonumber \\
	&\quad + \text{h.c.}   \nonumber
	\end{align} }
where as usual we defined the phonon displacement field $\phi_{\vec q,\mu} = a_{\vec q,\mu}+a_{-\vec q,\mu}^\dagger$\footnote{which is often defined with an additional factor $1/\sqrt{2}$}. In order to derive the $I-U$ characteristic of the system we have to apply a finite voltage $e U = \mu_S- \mu_T$, which can be done easily by the substitution $c_{\vec p,\sigma}(t) \rightarrow e^{-i \mu_T t} c_{\vec p,\sigma}(t)$ and $c_{\vec k,\sigma}(t) \rightarrow e^{-i \mu_S t} c_{\vec k,\sigma}(t)$. The dispersion energies of the tip $\epsilon_{\vec p}^T \rightarrow \xi_{\vec p}^T= \epsilon_{\vec p}^T- \mu_T$ and superconductor $\epsilon_{\vec k}^S \rightarrow \xi_{\vec k}^S= \epsilon_{\vec k}^S- \mu_T$ are now measured relative to their chemical potentials and this leads to a time dependence  of the tunneling matrix elements $T^{\text e} \rightarrow T^{\text e} e^{i e U t}, T^{\text{i}} \rightarrow T^{\text{i}} e^{i e U t}$ in the tunneling part $S_{\text{t}}$ of the action. 

\begin{figure}
	\centering
	\includegraphics[width=0.4\textwidth]{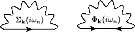}
	\caption{Normal and anomalous self-energy due to electron-phonon interaction that appear in the Eliashberg-theory.}
	\label{selfenergy}
\end{figure}

We build up our perturbation theory by rewriting $\erw{\ldots} = \int D[\bar c, c] D[\bar a,a] \ldots e^{i S} = \erw{e^{i S_{\text{t}}} \ldots }_0$ with the unperturbed expectation value $\erw{\ldots}_0 = \int D[\bar c, c] D[\bar a,a] \ldots e^{i S_0}$. The corresponding unperturbed propagators are then given in the $R,A,K$  basis (also known as \textit{Larkin-Ovchinnikov Representation})
\begin{align}
\hat G_{\vec k/\vec p}(t,t') &=-i\erw{\mx{c_{\vec k/\vec p,\sigma}^1(t)  \\ c_{\vec k/\vec p,\sigma}^2(t)  }\mx{c_{\vec k/\vec p,\sigma}^1(t')  \\ c_{\vec k/\vec p,\sigma}^2(t')  }^\dagger }_0  \nonumber \\
&= \mx{ G_{\vec k/\vec p}^R(t,t')& G_{\vec k/\vec p}^K(t,t') \\ 0 & G_{\vec k/\vec p}^A(t,t')}    \nonumber \\
\hat D_{\vec q,\mu}(t,t') &= -i \erw{\mx{\phi_{\vec q,\mu}^{\text{cl}} \\ \phi_{\vec q,\mu}^{\text{q}}} \mx{\phi_{-\vec q,\mu}^{\text{cl}} \\ \phi_{-\vec q,\mu}^{\text{q}}}^\dagger }_0  \nonumber \\
&= \mx{ D_{\vec q,\mu}^K(t,t') & D_{\vec q,\mu}^R(t,t') \\ D_{\vec q,\mu}^A(t,t') & 0}   \label{th6}
\end{align}
where the retarded propagators are given in energy representation $f(\omega)= \intinfty dt f(t) e^{i \omega t}$ as
\begin{align}
G_{\vec k}^{R}(\omega) &= \begin{matrix}
\includegraphics[width=8mm]{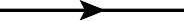}
\end{matrix} = \frac{Z_{\vec k}^{R}(\omega) \, \omega  + \xi_{\vec k}^S}{\bigl[Z_{\vec k}^{R}(\omega) \, \omega \bigr]^2- \bigl[\xi_{\vec k}^S \bigr]^2 - \Phi_{\vec k}^{R}(\omega)^2}   \nonumber \\
F_{\vec k}^{R}(\omega) &= \begin{matrix}
\includegraphics[width=8mm]{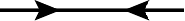}
\end{matrix} =\frac{- \Phi_{\vec k}^{R}(\omega)}{\bigl[Z_{\vec k}^{R}(\omega) \, \omega \bigr]^2- \bigl[\xi_{\vec k}^S \bigr]^2 - \Phi_{\vec k}^{R}(\omega)^2} \nonumber \\
G_{\vec p}^{R}(\omega) &=\begin{matrix}
\includegraphics[width=8mm]{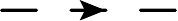}
\end{matrix} = \frac{1}{\omega- \xi_{\vec p}^S + i 0}    \nonumber \\
D_{\vec q,\mu}^{R}(\omega) &=\begin{matrix}
\includegraphics[width=8mm]{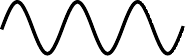}
\end{matrix} = \frac{2\omega_{\vec q,\mu}}{(\omega + i0)^2 - \omega_{\vec q,\mu}^2} \label{th7}
\end{align}
For the superconductor, we use the known framework of Eliashberg theory. Therefore, $Z_{\vec k}^R= 1 - \frac{\Sigma_{\vec k}^R(\omega)- \Sigma_{-\vec k}^A(-\omega)}{2\omega}$ is the renormalization function of the lead superconductor $S$ and $\Sigma_{\vec k}^{R/A}(\omega) = \Sigma_{\vec k}(\omega \pm i 0)$ is the phonon-induced normal self-energy, see Fig.~\ref{selfenergy}. We neglected the correction of the pure dispersion $\xi_{\vec k}^S$ due to the coupling to the phonons, because it will basically just give an unimportant shift of the chemical potential and can be assumed to be incorporated in the electronic dispersion already from the beginning. The anomalous self-energy $\Phi_{\vec k}^R(\omega)=\Phi_{\vec k}(\omega+i0)$ is depicted in Fig.~\ref{selfenergy} and we also gave the expression for the anomalous propagator $ F_{\vec k}^{R}(\omega)$, even tough we will only need the normal particle propagator (since in the NIS-junction there is no Josephson effect). We neglect the renormalization of the phonon spectrum due to the interaction with the electrons, which could be incorporated easily by a phonon self-energy that would just lead to a small broadening and modification of the phonon spectral weight. The effect of the Coulomb interaction between the fermions is as usual incorporated using a Coulomb pseudopotential $\mu^*$.
Since we consider the sub-systems $S$ and $T$ to be in thermal equilibrium, the Keldysh propagators have the simple structure
\begin{align}
G_{\vec k/\vec p}^{K}(\omega) &= \bigl[1-2 n_F(\omega) \bigr] \, \underbrace{\bigl[ G_{\vec k/\vec p}^{R}(\omega)-G_{\vec k/\vec p}^{A}(\omega)  \bigr]}_{- 2 \pi i A_{\vec k/\vec p}(\omega)}   \, , \nonumber  \\
D_{\vec q,\mu}^{K}(\omega)&= \bigl[ 1+ 2 n_B(\omega)\bigr]  \, \underbrace{\bigl[ D_{\vec q,\mu}^{R}(\omega)- D_{\vec q,\mu}^{A}(\omega) \bigr] }_{- 2 \pi i \bigl[ A_{\vec q, \mu}(\omega) - A_{\vec q, \mu}(-\omega) \bigr] } \, , \label{th8}
\end{align}
where $n_F(\omega)$ is the Fermi and $n_B(\omega)$ the Bose function with temperature $T$ and we defined the spectral weights $A_{\vec k/\vec p}(\omega),A_{\vec q,\mu}(\omega)$ of the electron and phonon systems. In our case $A_{\vec q,\mu}(\omega) = \delta(\omega-\omega_{\vec q, \mu})$. For completeness, we also give the explicit expressions for the greater/lesser Green functions
\begin{align}
G_{\vec k/\vec p}^{>/<}(\omega) &= - 2 \pi i \left\{  \begin{matrix}
1-n_F(\omega) \\ -n_F(\omega) \end{matrix}  \right\}  A_{\vec k/\vec p}(\omega)  \label{th8a} \\
D_{\vec q,\mu}^{>/<}(\omega)&= - 2 \pi i \left\{  \begin{matrix}
1+n_B(\omega) \\ n_B(\omega) \end{matrix}  \right\}  \biggl[ A_{\vec q, \mu}(\omega) - A_{\vec q, \mu}(-\omega) \biggr]    \nonumber 
\end{align}
Following Eq.~\zref{th4} it is easy to determine the explicit expression for the current
{ \small
	\begin{align}
	I &= i e \erw{\sum_{\vec k,\vec p} \mx{c_{\vec p,\sigma}^-(t) \\ c_{\vec k,\sigma}^-(t)}^\dagger  \mx{0 & -[T_{\vec k,\vec p}^+(t)]^* \\ T_{\vec k,\vec p}^-(t) &0 }  \mx{c_{\vec p,\sigma}^+(t) \\ c_{\vec k,\sigma}^+(t)} }     \nonumber \\
	&=  i e \erw{\sum_{\vec k,\vec p} \mx{c_{\vec p,\sigma}^-(t) \\ c_{\vec k,\sigma}^-(t)}^\dagger  \mx{0 & -[T_{\vec k,\vec p}^+(t)]^* \\ T_{\vec k,\vec p}^-(t) &0 }  \mx{c_{\vec p,\sigma}^+(t) \\ c_{\vec k,\sigma}^+(t)}  e^{i S_{\text t}}}_0     \nonumber \\
	&\approx - e \erw{\sum_{\vec k,\vec p} \mx{c_{\vec p,\sigma}^-(t) \\ c_{\vec k,\sigma}^-(t)}^\dagger  \mx{0 & -[T_{\vec k,\vec p}^+(t)]^* \\ T_{\vec k,\vec p}^-(t) &0 }  \mx{c_{\vec p,\sigma}^+(t) \\ c_{\vec k,\sigma}^+(t)}  S_{\text t}}_0  \label{th9}
	\end{align} } 
where we defined the total tunneling matrix element $T_{\vec k,\vec p}^\pm(t) = e^{i e U t} \bigl[T_{\vec k,\vec p}^{\text{e}}+ \sum_{\vec q,\mu} T_{\vec k,\vec p}^{\text{e}} \alpha_{\vec q,\mu} \phi_{\vec q,\mu}^\pm(t) \bigr]$ (with phonon field $\phi^\pm$ on the upper/lower Keldysh contour) and in the end expanded in leading order of $T_{\vec k,\vec p}$. Also, we defined the creation operators to be defined on the lower Keldysh contour (index $-$) and the annihilation operators on the upper Keldysh contour (index $+$), since the electron first has to leave one side before it can tunnel through the barrier to the other side. This time ordering can be conveniently expressed in Keldysh theory and we also assumed the phonons (field $\phi^\pm$) to be excited/absorbed on the same contour as the electrons on the superconductor $S$. The tunneling action can be written in a similar way as
{\small
	\begin{align}
	S_{\text t} &= -\int_{\mathcal C} dt \sum_{\vec k,\vec p} \mx{c_{\vec p,\sigma}(t) \\ c_{\vec k,\sigma}(t)}^\dagger  \mx{0 & [T_{\vec k,\vec p}(t)]^* \\ T_{\vec k,\vec p}(t) &0 }  \mx{c_{\vec p,\sigma}(t) \\ c_{\vec k,\sigma}(t)}  \nonumber \\
	&= -\int \limits_{-\infty}^\infty dt \sum_{\vec k,\vec p \atop \alpha = \pm} \alpha \,  \mx{c_{\vec p,\sigma}^\alpha(t) \\ c_{\vec k,\sigma}^\alpha(t)}^\dagger  \mx{0 & [T_{\vec k,\vec p}^\alpha(t)]^* \\ T_{\vec k,\vec p}^\alpha(t) &0 }  \mx{c_{\vec p,\sigma}^\alpha(t) \\ c_{\vec k,\sigma}^\alpha(t)}    \label{th10}
	\end{align}}

\subsection{Elastic current}
\begin{figure}
	\centering
	\includegraphics[width=0.3\textwidth]{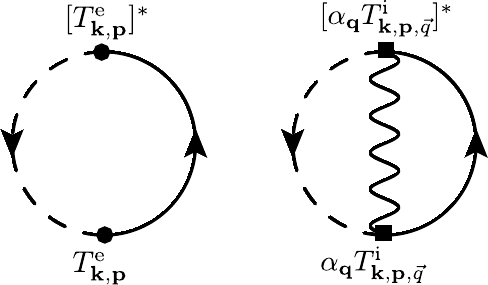}
	\caption{Feynman diagrams for the elastic (left) and inelastic (right) tunneling current in leading order $T^{\text e}, T^{\text i}$.}
	\label{Diagrams}
\end{figure}

Performing the contractions in Eq.~\zref{th9} we find the elastic current 
\begin{align}
I^{\text{e}}(U)&=  2 e \int \limits_{-\infty}^\infty  d\tau \, \frac{1}{V_S V_T} \sum_{\vec k, \vec p } \abs{T_{\vec k,\vec p}^{\text e}}^2 e^{i e U \tau}  \,  \sum_{ \alpha=\pm}\alpha \nonumber  \\
& \hspace{3mm} \bigl[G_{\vec k}^{\alpha,-}(-\tau)  G_{\vec p}^{+,\alpha}(\tau)  - G_{\vec k}^{+,\alpha}(-\tau)  G_{\vec p}^{\alpha,-}(\tau)  \bigr]    \nonumber  \\
&=  - 4 e \int \limits_{-\infty}^\infty  d\tau \, \frac{1}{V_S V_T}  \sum_{\vec k, \vec p } \abs{T_{\vec k,\vec p}^{\text e}}^2 e^{i e U \tau}   \label{th11} \\
& \hspace{3mm} \bigl[ \Im G_{\vec k}^<(-\tau)  \Im G_{\vec p}^R(\tau)  - \Im G_{\vec k}^R(-\tau)  \Im G_{\vec p}^<(\tau)  \bigr]     \nonumber
\end{align}
where we used the definition of the greater/lesser $G^{>/<}$ and time-ordered/anti-time-ordered  Green function $G^{\mathcal T / \tilde{\mathcal T}}$, see Section~\ref{secpropKeldysh} of the Supplementary Material. In the end, we used the known identities that relate $G^{\mathcal T / \tilde{\mathcal T}}$ to the greater, lesser, retarded and advanced propagators. In Fig.~\ref{Diagrams} we show the corresponding Feynman diagram for the elastic tunneling current for the leading order in the tunneling element. After transforming to Fourier space and inserting the explicit expressions of the propagators in thermal equilibrium, we find 
\begin{align}
I^{\text{e}}(U)&=   4 \pi e \int \limits_{-\infty}^\infty  d\omega \, \frac{1}{V_S V_T}  \sum_{\vec k, \vec p } \abs{T_{\vec k,\vec p}^{\text e}}^2   \label{th12} \\
& \hspace{3mm} \bigl[n_F(\omega)-n_F(\omega+eU) \bigr] A_{\vec k}(\omega) A_{\vec p}(\omega+eU)    \, ,  \nonumber
\end{align}
which is the usual expression for the elastic current in the Landauer-B\"uttinger transport theory assuming perfect quasiparticles $A_{\vec k/\vec p}(\omega)= \delta(\omega-\epsilon_{\vec k/\vec p}^{S/T})$. 

The elastic conductance $G^{\text e}(U)=\frac{dI^{\text e}}{dU} $  will now be calculated using the usual approximation $T_{\vec k, \vec p}^{\text e}= t^{\text e} = \text{const.}$ for small voltages $U \ll E_F \sim 1 \text{ eV}$, which is a reasonable assumption for an STM~\onlinecite{Giamarchi2011}. Assuming the DOS of the tip system $\nu^T(\omega)= 1/V_T \sum_{\vec p} A_{\vec p}(\omega) \approx \nu_T^0$ to be constant near the Fermi surface, we can then rewrite the elastic current \zref{th12} to be
\begin{align}
I^{\text{e}}(U)&=   4 \pi \nu_T^0 e \abs{t^{\text e}}^2 \intinfty  d \omega \,  \bigl[n_F(\omega)-n_F(\omega+eU) \bigr] \nu_S(\omega)   \, ,   \label{con2}
\end{align}
where we defined as usual the DOS of the superconductor as $\nu(\omega) = 1/V_S \sum_{\vec k}  A_{\vec k}(\omega)$. As it is well known, the elastic differential conductance is then given by
\begin{align}
\frac{dI^{\text e}}{dU} &= - 4 \pi \nu_T^0 e^2 \abs{t^{\text e}}^2 \intinfty d\omega \,  n_F'(\omega+eU) \nu_S(\omega) \nonumber \\
&=- \sigma_0  \intinfty d\omega \,  n_F'(\omega+eU) \tilde \nu_S(\omega) \label{con3a}
\end{align}
where we defined as the normalized DOS $\tilde \nu(\omega) = \nu(\omega)/\nu_S^0$ with $\nu_S^0$ as the DOS and $\sigma_0 =4 \pi \nu_T^0 \nu_S^0 e^2 \abs{t^{\text e}}^2$ as the elastic conductance in the normal state.  For small temperatures ($T \ll E_F$ in the normal conductor or $T \ll \Delta$ in the superconductor with gap $\Delta$), such that $n_F'(\epsilon) \approx -\delta(\epsilon)$, the elastic conductance simplifies to
\begin{align}
\frac{dI^{\text e}}{dU} &=  4 \pi \nu_T^0 e^2 \abs{t^{\text e}}^2  \nu(-e U) = \sigma_0 \tilde \nu(-eU)  \, ,\label{con3}
\end{align}
and is then proportional to the normalized DOS $\tilde \nu(\omega)$ of the superconductor. The corresponding expression von $d^2 I^{\text e} /dU^2$ can be computed easily from the expression~\zref{con3a}.

\subsection{Inelastic current}
The inelastic current can similarly be expressed by performing the contractions of~\zref{th9} containing the phonon-fields and expressing the occurring propagators in terms of retarded, advanced, greater and lesser propagators

\begin{widetext}
\begin{align}
I^{\text{i}}(U)&=  2 i e \int \limits_{-\infty}^\infty  d\tau \, \frac{1}{V_S^2 V_T}\sum_{\vec k, \vec p \atop \vec q , \mu} \abs{T_{\vec k,\vec p,\vec q, \mu}^{\text i} \alpha_{\vec q,\mu}}^2 e^{i e U \tau}   \sum_{ \alpha=\pm}\alpha \biggl[ G_{\vec k}^{\alpha,-}(-\tau) G_{\vec p}^{+,\alpha}(\tau)D_{\vec q,\mu}^{-,\alpha}(\tau)  - G_{\vec k}^{+,\alpha}(-\tau)G_{\vec p}^{\alpha,-}(\tau)   D_{\vec q,\mu}^{\alpha,+}(\tau) \biggr]  \nonumber  \\
&=  4 e \int \limits_{-\infty}^\infty  d\tau \, \frac{1}{V_S^2 V_T} \sum_{\vec k, \vec p \atop \vec q , \mu} \abs{T_{\vec k,\vec p,\vec q, \mu}^{\text i} \alpha_{\vec q,\mu}}^2 e^{i e U \tau}   \biggl[ \Im G_{\vec k}^{<}(-\tau) \Im G_{\vec p}^{<}(\tau) \Im D_{\vec q,\mu}^{R}(\tau) + \Im G_{\vec k}^{<}(-\tau) \Im G_{\vec p}^{R}(\tau) \Im D_{\vec q,\mu}^{>}(\tau)  \nonumber \\
&  \hspace{70mm}  - \Im G_{\vec k}^{R}(-\tau) \Im G_{\vec p}^{<}(\tau) \Im D_{\vec q,\mu}^{<}(\tau)   \biggr]      \label{ic1}
\end{align}
After going to Fourier space and inserting the corresponding electron and phonon propagators defined in Sec.~\ref{secpertapproach}, we can finally  rewrite the inelastic current as

	{\small
		\begin{align}
		I^{\text{i}}(U)&= - 4 \pi e \int d\omega_1 d\omega_2 \frac{1}{V_S^2 V_T} \sum_{\vec k,\vec p,\vec q \atop \mu} \abs{T_{\vec k,\vec p,\vec q, \mu}^i \alpha_{\vec q,\mu}}^2       \label{ic2}  \\
		& \hspace{3mm} \biggl[ A_{\vec q,\mu}(\omega_1) A_{\vec k}(\omega_2) A_{\vec p}(\omega_2-\omega_1+eU) \biggl( n_F(\omega_2-\omega_1+eU) n_B(\omega_1) \bigl[1-n_F(\omega_2) \bigr] - n_F(\omega_2) \bigl[1+n_B(\omega_1) \bigr] \bigl[1-n_F(\omega_2-\omega_1+eU) \bigr] \biggr)   \nonumber \\
		&\hspace{3mm} +A_{\vec q,\mu}(\omega_1)  A_{\vec k}(\omega_2) A_{\vec p}(\omega_2+\omega_1+eU)  \biggl( n_F(\omega_2+\omega_1+eU) \bigl[1+n_B(\omega_1)\bigr] \bigl[1-n_F(\omega_2) \bigr] - n_F(\omega_2) n_B(\omega_1)  \bigl[1-n_F(\omega_2+\omega_1+eU) \bigr] \biggr)   \biggr]  \nonumber
		\end{align}}
\end{widetext}

\begin{figure}
	\centering
	\includegraphics[width=0.45\textwidth]{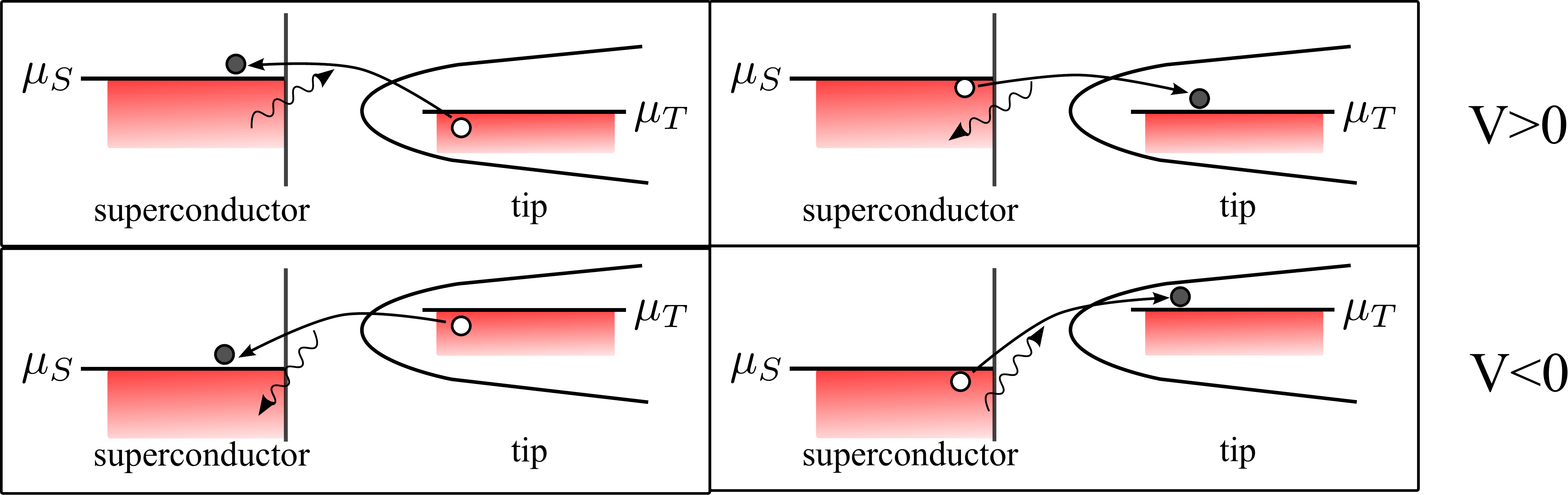}
	\caption{Inelastic tunneling processes for $V>0$ and $V<0$ via the emission/absorption of phonons.}
	\label{inelasticprocesses}
\end{figure}

The first/third term describes the tunneling of an electron from the tip to the superconductor via the absorption/excitation of a phonon and the second/fourth term the tunneling from the superconductor to the tip via a phonon excitation/absorption, see also Fig.~\ref{inelasticprocesses}. As in the elastic case, we apply the following simplifications: A constant inelastic vertex $T_{\vec k,\vec p,\vec q, \mu}^{\text i} = t^i$ and a constant DOS of the tip.  Let us define the weighted DOS of the phonons in the superconductor as
\begin{align}
\alpha^2   F_{\text{tun}}(\omega) &= \frac{1}{V_S}\sum_{\vec q,\mu} \abs{\alpha_{\vec q,\mu}}^2 A_{\vec q,\mu}(\omega)   \nonumber \\
&= \frac{1}{V_S} \sum_{\vec q,\mu} \abs{\alpha_{\vec q,\mu}}^2 \delta(\omega-\omega_{\vec q,\mu}) \, ,     \label{ic3}      
\end{align}
which is very similar to the Eliashberg function besides a different momentum average. For our case of very low temperature $k_B T \ll \omega_D$  only the processes that excite a phonon lead relevant inelastic contributions to the tunneling current since the number of thermal low-energy phonons $\alpha^2 F_{\text{tun}}(\omega) \cdot n_B(\omega) \approx 0 $ in the system is negligible.  We then find 
\begin{align}
I^{\text i} &= 4 \pi e \nu_T^0 \nu_S^0\abs{t^i}^2 \int d\omega_1 d\omega_2 \alpha^2   F_{\text{tun}}(\omega_1)  \tilde \nu_S(\omega_2)  \label{ic4}     \\
& \quad \biggl(n_F(\omega_2) n_F(\omega_1-\omega_2-eU)   - \{\omega_2,U \rightarrow - \omega_2,-U \}    \biggr)   \nonumber 
\end{align}
For the differential conductance we then find
\begin{align}
\frac{dI^\text{i}}{dU} &= -\sigma_0 \abs{\frac{t^{\text i}}{t^{\text e}}}^2\int d\omega_1 d\omega_2 \alpha^2   F_{\text{tun}}(\omega_1)  \tilde \nu_S(\omega_2)  \label{ic5}   \\
&  \biggl(n_F(\omega_2) n_F'(\omega_1-\omega_2-eU) + \{\omega_2,U \rightarrow - \omega_2,-U \}    \biggr)  \nonumber \\
&= \sigma_0 \abs{\frac{t^{\text i}}{t^{\text e}}}^2\int d\omega   \biggl[ \alpha^2   F_{\text{tun}}^{T}(\omega+eU)  \tilde \nu_S(\omega) n_F(\omega)     \nonumber \\
& \hspace{1.8cm} + \alpha^2   F_{\text{tun}}^{T}(\omega-eU)  \tilde \nu_S(-\omega)n_F(\omega)  \biggr]
\end{align} 
where we defined the thermal broadened weighted DOS of the phonons as the convolution (in the limit of zero temperature it obviously holds $\alpha^2 F_{\text{tun}}^{T=0}(\omega) = F_{\text{tun}}(\omega)$)
\begin{align}
\alpha^2 F_{\text{tun}}^T(x) = - \intinfty dy \,  \alpha^2 F_{\text{tun}}(y) n_F'(y-x)   \label{ic6}
\end{align}
For particle-hole symmetric electronic systems $\tilde \nu_S(\omega) = - \tilde \nu_S(-\omega)$ and in the limit that $k_B T$ is much smaller than the characteristic phonon frequencies (meaning  $F_{\text{tun}}^{T}(\omega\pm eU) n_F(\omega) \simeq \theta(\mp U)$) , we can simplify this expression to
{\small{\begin{align}
		\frac{dI^\text{i}}{dU} &= \sigma_0 \abs{\frac{t^{\text i}}{t^{\text e}}}^2\int d\omega   \,  \alpha^2   F_{\text{tun}}^{T}(\omega+e \abs{U}) \tilde \nu_S(\omega) n_F(\omega)        \label{ic7}
		\end{align} }}

\section{Elastic and inelastic STM tunneling for single mode}
In order to get a qualitative  understanding of the inelastic tunneling contribution in the superconducting state, let us analyze a simple toy model. The toy model consists of a single phonon mode with $\alpha^2 F_{\text{tun}}(\omega) \simeq \alpha^2 F(\omega)= A_0 f(\omega) \frac{\gamma_0}{(\omega-\omega_0)^2+\gamma_0^2} $  with characteristic phonon energy $\omega_0=5 \meV$  and  half-width $\gamma_0$ . The function
$f(\omega)=\frac{\omega^2}{\omega^2+(1\meV)^2}$ ensures the proper low frequency behavior of acoustic phonons and rapidly approaches unity for larger frequencies. 
\begin{figure}
	\centering
	\includegraphics[width=0.4\textwidth]{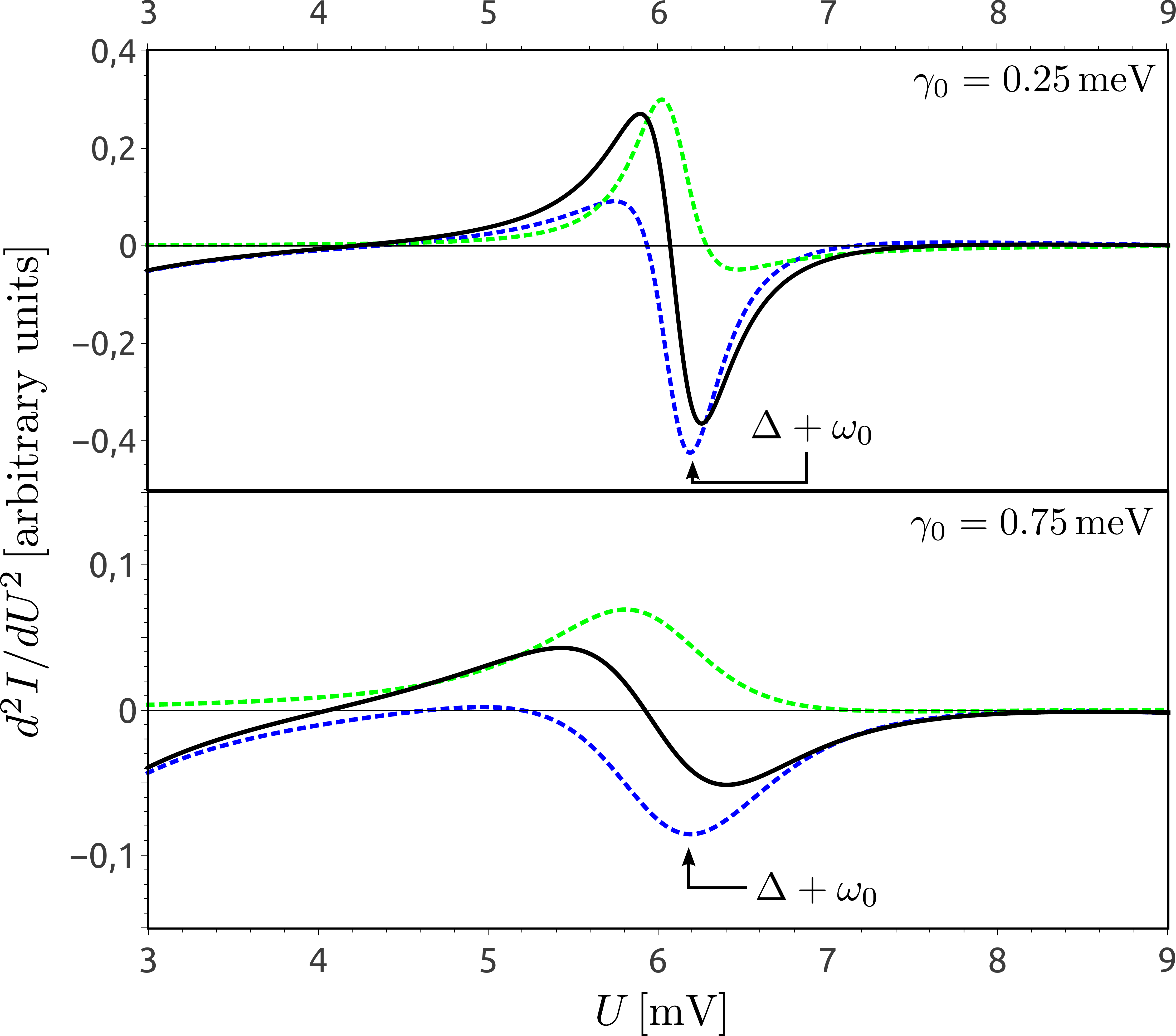}
	\caption{Elastic (blue dashed), inelastic (green dashed) contribution to the total (black solid) second derivative of the current $d^2 I /d U^2$ in the superconducting state for different peak width $\gamma_0$. The additional inelastic contribution will lead to a peak-dip feature with similar positive and negative amplitude in the second derivative, whereas the purely elastic contribution only pronounces the dip at the shifted phonon frequency $eU \approx \Delta+\omega_0$ strongly.}
	\label{SingleMode}
\end{figure}
The value $A_0$ is choosen such that the dimensionless electron-phonon coupling constant $\lambda=2 \int_0^\infty d\omega \, \alpha^2 F(\omega)/\omega = 1.5$. We use $\mu^*=0.1$ for the pseudopotential, such that solving the Eliashberg equations yields a gap value $\Delta \simeq 1 \meV$. In Fig.~\ref{SingleMode} we see the resulting second derivative of the tunneling current, which is more sensitive to the fine-structure than the conductance, for the above mentioned  elastic and inelastic tunneling for different peak width $\gamma_0$ in the superconducting state. As was seen in the experiments we use the ratio $\abs{t^{\text i}/t^{\text e}}^2 \approx 0.12 / \int_0^{10 \meV} d\omega \,  \alpha^2 F(\omega)$.
The inelastic contribution Eq.~\zref{ic5} has its major contribution for frequencies a bit below the energy of the phonon peaks shifted by the gap $\Delta$. Since inelastic tunneling adds additional channels to the conductance, its contribution will have the opposite sign of the elastic contribution in \zref{con3a} and can give pronounced positive peaks in the second derivative of the tunneling current followed by a negative peak of same amplitude . These symmetric peak-dip features around the zero axis in the second derivative of the tunneling current are characteristic for the joint elastic and inelastic STM. In the present example, we were not able to find an Eliashberg function that would yield such a tunneling spectra from the purely elastic tunneling contribution Eq.~\zref{con3a}. Even for very sharp Eliashberg spectra where the second derivative of the tunneling current can be clearly positive for some voltages, see upper picture in Fig.~\ref{SingleMode}, the following dip will always be much more pronounced if one only considers the elastic tunneling contribution. 

\section{Modelling of experimental broadening}
The experimental resolution is limited due to the used lock-in technique. As was shown in Ref.~\onlinecite{Klein73}, for a modulation voltage $V_{\text{mod}}$ (the modulation voltages stated in this work are root mean square values $U_{\text{mod}} = U_{\text{mod,max}}/\sqrt{2}$) the experimental curve for the second derivative measurements is the actual current convoluted with a Gaussian function $\Gamma(\omega)$ of half-width FWHM$ \approx 1.7 e V_{\text{mod}}$. This corresponds in our case to a standard deviation $\sigma=\text{FWHM}/2.4 =0.31 \text{meV}$. For the elastic part this means that we get
\begin{align}
\frac{d^2I^{\text{e,exp}}}{dU^2} &= - \sigma_0 \intinfty dE  d\omega  \, \Gamma(eU-E)    n_F'(\omega+E) \tilde \nu_S(\omega)    \label{exp1}
\end{align}
which can be easily computed using the fermionic DOS obtained from solving the Eliashberg equations. 

Let us now consider the experimental broadening for inelastic tunneling and restrict us to the case of positive bias voltages. Following Eq.~\zref{ic7}, the second derivative of the inelastic tunneling current is for positive $U>0$ given by
\begin{align}
\frac{d^2I^{\text{i}}}{dU^2} &=  e \sigma_0 \abs{\frac{t^{\text i}}{t^{\text e}}}^2  \intinfty  d\omega   \,  \alpha^2   F_{\text{tun}}^{T}{}'(eU+\omega) \tilde \nu_S (\omega) n_F(\omega)    \label{exp2}   
\end{align}
To get the  experimental data (both in the normal and superconducting state), we have to broaden the function by the convolution
\begin{align}
\frac{d^2I^{\text{i,exp}}}{dU^2} &=  e \sigma_0 \abs{\frac{t^{\text i}}{t^{\text e}}}^2  \intinfty dE    \, \Gamma(eU-E)  \label{exp3} \\
& \hspace{1.3cm} \intinfty d\omega \, \alpha^2   F_{\text{tun}}^{T}{}'(E+\omega) \tilde \nu_S (\omega) n_F(\omega)    \nonumber \\
&=  e \sigma_0 \abs{\frac{t^{\text i}}{t^{\text e}}}^2   \int d\omega \, \alpha^2   F_{\text{tun}}^{\text{exp}}{}'(eU+\omega) \tilde \nu_S (\omega)   n_F(\omega)  \nonumber
\end{align}
with the thermally and modulation voltage broadened spectral function
\begin{align}
\alpha^2   F_{\text{tun}}^{\text{exp}}(x) = \intinfty dy \Gamma(x-y)\alpha^2   F_{\text{tun}}^{T}(y)   \label{exp4}
\end{align}
In the normal state $\tilde \nu_S(\omega) \approx 1$ this simplifies in the limit of small temperatures $T \ll \omega_D, E_F$ (such that $n_F(\omega) \approx \theta(-\omega)$) to
\begin{align}
\frac{d^2I_{\text{nc}}^{\text{i,exp}}}{dU^2} &\approx e \sigma_0 \abs{\frac{t^{\text i}}{t^{\text e}}}^2   \intinfty d\omega \, \alpha^2   F_{\text{tun}}^{\text{exp}}{}'(eU+\omega)    n_F(\omega)   \nonumber \\
&= e \sigma_0 \abs{\frac{t^{\text i}}{t^{\text e}}}^2 \alpha^2   F_{\text{tun}}^{\text{exp}}(eU)   \label{exp5}
\end{align}
As $\frac{d^2I_{\text{nc}}^{\text{e,exp}}}{dU^2} \approx 0$, we can extract the $\alpha^2   F_{\text{tun}}^{T,\text{mod}}(eU)$ function from the normal state measurements and can then use it to calculate the inelastic current in the superconducting state.

\section{Multiple phonon processes}
If we consider tunneling processes with a higher number of excited phonons we can formally write down the following tunneling processes
\begin{align}
\delta H_{\text t}^{(n)} &=  \frac{1}{\sqrt{V_t} V_S^{\frac{n+1}{2}}}\sum_{\vec k,\vec p, \vec q_1, \ldots, \vec q_n \atop \sigma, \mu_1, \ldots , \mu_n} T_{\vec k,\vec p, \vec q_1, \ldots, \vec q_n, \mu_1, \ldots, \mu_n}^{\text i}     \nonumber \\
& \quad \alpha_{\vec q_1, \mu_1} \ldots \alpha_{\vec q_n,\mu_n} c_{\vec k,\sigma}^\dagger c_{\vec p,\sigma}   \phi_{\vec q_1, \mu_1} \ldots \phi_{\vec q_n,\mu_n}    \label{ic16}
\end{align}
In the zero temperature limit it is then straightforward to generalize the result \zref{ic4} to the n-phonon process (demanding energy conservation and Fermi statistic for the leads)
\begin{widetext}
	\begin{align}
	I^{\text{i},(n)} &=  4 \pi e \nu_T^0 \abs{t^{\text i,(n)}}^2 \int d\omega_1 \ldots d\omega_{n} d\omega_{n+1} \alpha^2    F_{\text{tun}}(\omega_1) \ldots \alpha^2    F_{\text{tun}}(\omega_n)  \nu_S(\omega_{n+1})   \label{ic17}        \\
	& \hspace{2mm} \biggl[  \theta(-\omega_{n+1}) \theta(\omega_{n+1}- \omega_{n}-\ldots - \omega_1+ e U) - \{\omega_{n+1},U \rightarrow - \omega_{n+1},-U \} \biggr]\nonumber 
	\end{align}
	For the conductance, we then find for particle-hole symmetric systems
	\begin{align}
	\frac{d I^{\text{i},(n)} }{dU} &= \sigma_0 \abs{\frac{t^{\text i,(n)}}{t^{\text e}}}^2 \text{sign}(U)  \int_0^\infty d\omega  d\omega_1  \ldots d\omega_{n-1} \alpha^2  F_{\text{tun}}(e \abs{U}-\omega-\omega_1) \alpha^2  F_{\text{tun}}(\omega_1-\omega_2) \ldots \alpha^2  F_{\text{tun}}(\omega_{n-2}-\omega_{n-1}) \tilde \nu_S(\omega)    \nonumber \\
	&=  \sigma_0 \abs{\frac{t^{\text i,(n)}}{t^{\text e}}}^2 \text{sign}(U)  \int_0^\infty d\omega \,  \alpha^{2n}  F_{\text{tun}}^n(e \abs{U}-\omega) \tilde \nu_S(\omega)    \label{ic18}
	\end{align}
	where we defined the convolution
	\begin{align}
	\alpha^{2n}  F_{\text{tun}}^n(\omega) &= \int_0^\infty d\omega_1 \ldots d\omega_{n-1} \,  \alpha^2  F_{\text{tun}}(\omega-\omega_1) \alpha^2  F_{\text{tun}}(\omega_1-\omega_2) \ldots \alpha^2  F_{\text{tun}}(\omega_{n-2}-\omega_{n-1})   \label{ic19}
	\end{align}

\end{widetext}

\section{Important relations of Non-Equilibrium propagators}  \label{secpropKeldysh}
Following Ref.~[\onlinecite{Keldyshbook}] we define for both the fermionic and the bosonic fields $\phi(t)$ the greater, lesser, time-ordered and anti-time-ordered Green's functions as
\begin{align}
\begin{split}
G^<(t,t') &= G^{+-}(t,t') =- i \erw{\phi^+(t) \bar \phi^-(t')}   \\
G^>(t,t') &= G^{-+}(t,t') =- i \erw{\phi^-(t) \bar \phi^+(t')}   \\
G^{\mathcal T } (t,t') &= G^{++}(t,t') = - i \erw{\phi^+(t) \bar \phi^+(t')}   \\
G^{\tilde{\mathcal T} } (t,t') &= G^{--}(t,t') = - i \erw{\phi^-(t) \bar \phi^-(t')}  
\end{split}    \label{pK1}
\end{align}
We can now perform the Keldysh rotation to the classical and quantum fields in the bosonic case:
\begin{align}
\begin{split}
\phi^{\text{cl}}(t) &= \frac{1}{\sqrt{2}} \bigl[\phi^+(t)+ \phi^-(t) \bigr]  \\
\phi^{\text{q}}(t) &= \frac{1}{\sqrt{2}} \bigl[\phi^+(t)- \phi^-(t) \bigr]
\end{split}    \label{pK2}
\end{align}
and similar for the conjugate fields $\bar \phi(t)$. However, for the fermionic fields we use the Ovchinnikov-Larkin
convention~[\onlinecite{Larkin}]
\begin{align}
\begin{split}
\phi^{\text{1}}(t) &= \frac{1}{\sqrt{2}} \bigl[\phi^+(t)+ \phi^-(t) \bigr]  \\
\phi^{\text{2}}(t) &= \frac{1}{\sqrt{2}} \bigl[\phi^+(t)- \phi^-(t) \bigr]  \\
\bar \phi^{\text{1}}(t) &= \frac{1}{\sqrt{2}} \bigl[\bar \phi^+(t)- \bar \phi^-(t) \bigr]  \\
\bar \phi^{\text{2}}(t) &= \frac{1}{\sqrt{2}} \bigl[\bar \phi^+(t)+ \bar \phi^-(t) \bigr]
\end{split}    \label{pK3}
\end{align}
For the fermionic and bosonic cases, the retarded, advanced and Keldysh propagators are then defined as in Eq.~\zref{th6}. The relations between the different Green's functions ($>,<,\mathcal T, \tilde{\mathcal T}, R,A,K$) can be summarized by
\begin{align}
\begin{split}
0 &=G^{\mathcal T} + G^{\tilde{\mathcal T}} - G^>-G^<   \\
G^K &= G^>+G^< \\
G^R &= \frac{1}{2} \bigl[ G^{\mathcal T} - G^{\tilde{\mathcal T}}  +  G^> - G^<  \bigr]    \\
G^A &= \frac{1}{2} \bigl[ G^{\mathcal T} - G^{\tilde{\mathcal T}}  -  G^> + G^<  \bigr]
\end{split}   \label{pK4}
\end{align}

%\begin{acknowledgments}%The authors are grateful to ?????? for discussions. This work was supported by ...\textbf{nur experimenteller Support!!!}.%\end{acknowledgments}

\end{document}